\documentclass[12pt]{article}
\usepackage{graphicx}
\usepackage{amsmath}
\usepackage{amssymb}
\usepackage{caption2}
\begin{document}
\bibliographystyle{prsty}
\begin{center}
{\large {\bf \sc{   A new approach for calculating the Nambu-Gorkov propagator in color superconductivity theory }}} \\[2mm]
Z. G. Wang$^{1}$ \footnote{Corresponding author; E-mail,wangzgyiti@yahoo.com.cn.  }, S. L. Wan$^{2}$ and W. M. Yang$^{2} $    \\
$^{1}$ Department of Physics, North China Electric Power University, Baoding 071003, P. R. China \\
$^{2}$ Department of Modern Physics, University of Science and Technology of China, Hefei 230026, P. R. China \\
\end{center}

\begin{abstract}
 In this article, we propose a new approach to calculate  the
Nambu-Gorkov propagator intuitively with some linear algebra
techniques in presence of the scalar diquark condensates. With the
help of energy projective operators, we can obtain relatively simple
expressions for the quark propagators, which greatly facilitate the
calculations  in solving the Schwinger-Dyson equation to obtain the
gap parameters.
\end{abstract}

PACS : 12.38.-t, 12.38.Aw, 26.60.+c

{\bf{Key Words:}}  Diquark condensate, color superconductivity
\section{Introduction}

At  sufficiently low temperature and high density, the hadrons may
be crushed into quark matter, which may exist in the central region
of compact stars, where the densities are above the nuclear density
and  the temperatures are of the order of tens of {\rm keV}
\cite{AlfordRapp, ReviewCD}. The color superconducting quark matter
may also be created in future low-energy heavy ion colliders. The
essence of color superconductivity is the pairing between the quarks
when there exists an attractive interaction at the Fermi surface,
just like the BCS mechanism in QED.  Perturbatively, the quark-quark
interaction is dominated by the one-gluon exchange, which is simply
the one-photon exchange in QED multiplying the group  factor
$T^A_{ij}T_A^{kl}$, the group factor  can be decomposed into an
anti-triplet  $\bar{3}$ with attraction and a sextet $6$ with
repulsion,
\begin{equation}
 T^A_{ij}T^A_{kl}
=-{N_c+1\over 4N_c}(\delta_{il}\delta_{jk} -\delta_{ij}\delta_{kl})
+{N_c-1\over 4N_c}(\delta_{il}\delta_{jk} +\delta_{ij}\delta_{kl}).
\nonumber
\end{equation}
The quarks, unlike the electrons, have color and flavor as well as
spin degrees of freedom,  many different patterns of pairing are
possible, which lead to the copious phase structures in color
superconductivity theory,  such as  the color-flavor locking (CFL)
state \cite{CFL}, gapless color-flavor locking (gCFL) state
\cite{gCFL}, 2SC state, gapless 2SC (gS2C) state \cite{g2SC},  2SCus
state and Larkin-Ovchinnikov-Fudde-Ferrell (LOFF) state \cite{LOFF},
etc. Intense theoretical works  have been done to investigate the
different regions of the QCD phase diagram in past years
\cite{ReviewCD}.  The scaling formula relating the transition
temperature and the energy gap to the chemical potential and the
running coupling constant  have been derived systematically at
sufficiently high density,  which may serve a rigorous proof of the
existence of the color superconducting phase. At the region of
moderate baryon density or below,   it is too difficult to warrant a
first principle investigation with the present analytical and
numerical techniques. Furthermore, the charge neutrality condition
as well as the $\beta$ equilibrium have to be taken into account  to
form bulk matter inside the compact stars, which induce  mismatches
between the Fermi surfaces of the pairing quarks and puts severe
constraint in determining  the true ground states. We have to resort
to the effective actions, for example, the QCD-inspired
Nambu-Jona-Lasinio (NJL) like models \cite{NJL},  or solving the
Schwinger-Dyson equation (SDE) for the gaps consistently
\cite{ColorSD}. In solving the SDE, the manipulations in
color-flavor-spin spaces is tedious and cumbersome. In this article,
we propose a new approach to calculate the quark propagator in
presence of the scalar diquark condensates.

The article is arranged as follows:  we introduce the
Nambu-Jona-Lasinio model for illustrating  the formation of the
scalar  diquark condensate in section II; in section III,  we
propose a new approach to calculate the Nambu-Gorkov propagator
explicitly in presence of the scalar diquark condensates, which is
of special importance  in  solving the Schwinger-Dyson equation;
section IV is reserved for conclusion.

\section{Nambu-Jona-Lasinio model for diquark condensate}
At low temperature and  high ( or moderate) baryon density, the
quarks can pair with each other at their   Fermi surfaces with
sufficiently strong attractions and form Cooper pairs.  The colored
Cooper pairs ( represented by the field $\Delta^a$)  have
interactions with the quarks,  its vacuum expectation values ( the
diquark condensates ) induce the energy gaps for the quarks in the
superconducting phases. The formation of the diquark condensates can
not be derived from the first principle of QCD, we resort to the
QCD-inspired  NJL model or extended-NJL model to illustrate  this
subject \cite{NJL}. The NJL model or extended-NJL model have the
same symmetries as QCD and  can describe the spontaneous breakdown
of chiral symmetry in the vacuum and its restoration at high
temperature and high density.

For simplicity,  we can consider the two-flavor Lagrangian density
with $SU(2)_L \times SU(2)_R$ chiral symmetry, the extension to the
three-flavor NJL model with $SU(3)_L \times SU(3)_R$ chiral symmetry
is straightforward and easy,
\begin{eqnarray}
{\cal L} = {\bar q}(i\gamma^{\mu}\partial_{\mu}-m_0+\gamma^0 \mu )q
+
   G_S[({\bar q}q)^2 + ({\bar q}i\gamma_5{\bf {\bf \tau}}q)^2 ]
 +G_D[(i {\bar q}^C  \varepsilon  \epsilon^{b} \gamma_5 q )
   (i {\bar q} \varepsilon \epsilon^{b} \gamma_5 q^C)],
\end{eqnarray}
here  $q^C=C {\bar q}^T$, ${\bar q}^C=q^T C$ with  $C=i \gamma^2
\gamma^0$.  The $m_0$ is the current quark mass and the $\mu$ is the
chemical potential.  The quark field $q= q_{\alpha,i}$ with
$\alpha=1,2,3$ are  color index and  $i=1,2$ are flavor index, ${\bf
\tau}=(\tau^1,\tau^2,\tau^3)$ are Pauli matrixes,
$(\varepsilon)^{ik} = \varepsilon^{ik}$ and $(\epsilon^b)^{\alpha
\beta} = \epsilon^{\alpha \beta b}$ are totally antisymmetric
tensors in the flavor and color spaces respectively.

We can introduce the auxiliary fields and perform the standard
bosonization to obtain the following linearized  model in the
mean-field approximation,
\begin{eqnarray}
{\cal L} & =  & {\bar q}(i\gamma^{\mu}\partial_{\mu}-m_0+\gamma^0
\mu)q -
  {\bar q}(\sigma+i \gamma^5{\bf \tau}{\bf \pi}) q \nonumber\\
  &&-
  \frac{1}{2}\Delta^{*b} (i{\bar q}^C  \varepsilon \epsilon^{b}\gamma_5 q )
  -\frac{1}{2}\Delta^b (i {\bar q}  \varepsilon  \epsilon^{b} \gamma_5 q^C) +\cdots
  ,
 \end{eqnarray}
with the bosonic fields
\begin{eqnarray}
\Delta^b \sim i {\bar q}^C \varepsilon \epsilon^{b}\gamma_5 q \, , \
\ \Delta^{*b} \sim i {\bar q}  \varepsilon  \epsilon^{b} \gamma_5
q^C \, , \ \ \sigma \sim {\bar q} q \, , \ \ {\bf \pi} \sim i {\bar
q}\gamma^5 {\bf \tau} q \, .
\end{eqnarray}
The conditions $\langle \sigma \rangle \neq 0$  and $\Delta^b \neq
0$ indicate that the chiral symmetry and the color symmetry are
spontaneously broken respectively. Here only the red and green
quarks participate in the condensate, while the blue quarks (with
color index $b$) do not. We  can introduce the constituent quark
mass
\begin{eqnarray}
m=m_0+ \langle \sigma \rangle.
\end{eqnarray}
The spontaneously chiral symmetry breaking induced by the formation
of chiral condensate in the vacuum and the dynamical chiral symmetry
breaking induced by the strong color attraction at low energy region
can all result  in constituent masses for the quarks, for the $u$
and $d$ quarks, $m_{u,d} \approx 330 {\rm MeV}$, for the $s$ quark,
$m_s \approx 500 {\rm MeV}$. The constituent quark mass $m$ changes
according to the variations of the  baryon density, in this article,
we take the mass matrix for the $u$, $d$ and $s$ quarks to be
\begin{eqnarray}
    \left(
          \begin{array}{ccc}
             m  &  0&0 \\
            0  &  m&0\\
            0&0&m+\delta m \\
              \end{array}
             \right),
\end{eqnarray}
here the $m$ denotes  the degenerate mass and the the mass breaking
term $\delta m$ can be taken into account by modifying  the
corresponding chemical potential for the $s$ quark.

\section{Nambu-Gorkov propagator with diquark condensates}

We introduce the Nambu-Gorkov formulation  for the basis $q_{ru}$,
$q_{gu}$, $q_{rd}$,  $q_{gd}$,
\begin{eqnarray}
{\Psi}^T &=& (  q_{ru},  q_{gu},  q_{rd},  q_{gd};  q_{ru}^C,
q_{gu}^C, q_{rd}^C, q_{gd}^C ) \, , \nonumber \\
 {\bar \Psi}& =& ( {\bar
q}_{ru}, {\bar q}_{gu}, {\bar q}_{rd}, {\bar q}_{gd}; {\bar
q}_{ru}^C, {\bar q}_{gu}^C,{\bar q}_{rd}^C,{\bar q}_{gd}^C ) \, ,
\end{eqnarray}
and write down the inverse Nambu-Gorkov propagator in momentum
space,
\begin{equation}
{\rm G}^{-1} =
    \left(
          \begin{array}{cc}
             [G_0^{+}]^{-1}  &  \Delta^{-} \\
            \Delta^{+}  &  [ G_0^{-}]^{-1}
              \end{array}
             \right),
\end{equation}
with
\begin{eqnarray}
[G_0^{\pm}]^{-1}&=& \left(
          \begin{array}{cccc}
            [G_0^{\pm}]_{ru}^{-1}  &  0  & 0 & 0\\
            0 &  [G_0^{\pm}]_{gu}^{-1}  & 0 & 0\\
            0  & 0 & [G_0^{\pm}]_{rd}^{-1}  &  0 \\
            0 & 0  & 0 &  [G_0^{\pm}]_{gd}^{-1}
              \end{array}
             \right), \\
\Delta^{-} &=& -i \Delta \gamma_5 \left(
          \begin{array}{cccc}
            0  &  0  & 0 & 1\\
            0 & 0 & -1 & 0\\
            0  & -1 & 0  &  0 \\
            1 & 0  & 0 & 0
              \end{array}
             \right),  \  \ \\
\Delta^{+} &=& \gamma^0 (\Delta^{-})^{\dagger} \gamma^0 \,  , \\
 \left[G_0^{\pm}\right]_{\alpha i}^{-1}&=& (p_0 \pm \mu_{\alpha i}) \gamma_0
-{\mbox{\boldmath$\gamma$}}\cdot {\bf p} -m \, .
\end{eqnarray}
 At moderate baryon density, the quark masses  have
contribution from the chiral condensate and can not be neglected. We
have to use the following  energy projective operators  for massive
free particles \cite{Huang01}, which are used extensively  in
solving the Bethe-Salpeter equation \cite{Greiner},
\begin{eqnarray}
\Lambda_{\pm}({\vec p})&=&\frac{1}{2}(1\pm\frac{\gamma_0({\vec
\gamma}\cdot{\vec p}+m)}{E}) \ , \\
 \tilde \Lambda_{\pm}({\vec p})&=&
\frac{1}{2}(1\pm\frac{\gamma_0({\vec \gamma}\cdot{\vec p}-m)}{E}) \
, \
\end{eqnarray}
with the energy $E=\sqrt{{\vec p}^2+m^2}$ and the basic properties,
\begin{eqnarray}
 \Lambda_{\pm}({\vec p})\Lambda_{\pm}({\vec p})&=&\Lambda_{\pm}({\vec p}) \ , \nonumber \\
 \Lambda_{\pm}({\vec p})\Lambda_{\mp}({\vec p})&=&0 \ , \nonumber \\
 \Lambda_{+}({\vec p})+\Lambda_{-}({\vec p})&=&1 \ , \nonumber \\
 \gamma_0 \Lambda_{\pm}({\vec p}) \gamma_0&=&\tilde \Lambda_{\mp}({\vec p})\ , \nonumber \\
 \gamma_5 \Lambda_{\pm}({\vec p}) \gamma_5&=&\tilde
\Lambda_{\pm}({\vec p})\ . \nonumber
\end{eqnarray}
 With those projective operators, we  re-write the matrix elements for the normal quark propagator
 in the following form which can greatly facilitate the
calculations as the terms concerning the operators  $\Lambda_{+}$
and $\Lambda_{-}$ decouple from each other when inverting  the
matrix,
\begin{eqnarray}
 \left[G_0^{\pm}\right]^{-1}& = &
\gamma_0(p_0-E^{\mp})\Lambda_{+}+\gamma_0(p_0+E^{\pm})\Lambda_{-} \ ,  \\
G_0^{\pm}& = &  \frac{\gamma_0\tilde \Lambda_{+}}{p_0+E^{\pm}} +
\frac{\gamma_0\tilde\Lambda_{-}}{p_0-E^{\mp}} \ ,
\end{eqnarray}
here $E^\pm = E \pm \mu$.

In the following, we  invert  the matrix in Eq.(7) with some linear
algebra techniques to derive the Nambu-Gorkov propagator $G$
intuitively. For the CFL and gCFL states, the  $G^{-1}$ is a
$72\times72$ matrix in the color-flavor-spin spaces, it is very
difficult to obtain the Nambu-Gorkov propagator $G$ with simple
expressions, the present approach can greatly facilitate the
calculations,

\begin{eqnarray}
{\rm G} &=&\frac{1}{\left(
          \begin{array}{cc}
            \left[ G_0^{+} \right]^{-1}  &  \Delta^{-} \\
            \Delta^{+}  &  \left[ G_0^{-} \right]^{-1}
              \end{array}\right)} \, ,    \nonumber\\
             &=& \frac{1}{\left(
          \begin{array}{cc}
            \left[ G_0^{+} \right]^{-1}  & 0\\
            0  &  \left[ G_0^{-} \right]^{-1}
              \end{array}\right)+\left(
          \begin{array}{cc}
            0  &  \Delta^{-} \\
            \Delta^{+}  & 0
              \end{array} \right)} \, ,\nonumber\\
              &=& \frac{1}{\left(
          \begin{array}{cc}
            \left[ G_0^{+} \right]^{-1}  & 0\\
            0  &  \left[ G_0^{-} \right]^{-1}
              \end{array}\right)\left\{1+\left(\begin{array}{cc}
             G_0^{+}   & 0\\
            0  &   G_0^{-}
              \end{array}\right)\left(
          \begin{array}{cc}
            0  &  \Delta^{-} \\
            \Delta^{+}  & 0
              \end{array} \right) \right\}} \, , \nonumber\\
              &=& \frac{1}{\left(
          \begin{array}{cc}
            \left[ G_0^{+} \right]^{-1}  & 0\\
            0  &  \left[ G_0^{-} \right]^{-1}
              \end{array}\right)\left\{1+\left(
          \begin{array}{cc}
            0  &  G_0^{+}\Delta^{-} \\
            G_0^{-}\Delta^{+}  & 0
              \end{array} \right) \right\}}\, , \nonumber\\
              &=& \frac{1}{\left(
          \begin{array}{cc}
            \left[ G_0^{+} \right]^{-1}  & 0\\
            0  &  \left[ G_0^{-} \right]^{-1}
              \end{array}\right)}\left\{1-\left(
          \begin{array}{cc}
            0  &  G_0^{+}\Delta^{-} \\
            G_0^{-}\Delta^{+}  & 0
              \end{array} \right) +\left(
          \begin{array}{cc}
            G_0^{+}\Delta^{-}G_0^{-}\Delta^{+}  & 0  \\
             0 & G_0^{-}\Delta^{+}G_0^{+}\Delta^{-}
              \end{array} \right)\right. \,  \nonumber\\
              &&\left.- \left(
          \begin{array}{cc}
            0  &  G_0^{+}\Delta^{-} \\
            G_0^{-}\Delta^{+}  & 0
              \end{array} \right)\left(
          \begin{array}{cc}
            G_0^{+}\Delta^{-}G_0^{-}\Delta^{+}  & 0  \\
             0 & G_0^{-}\Delta^{+}G_0^{+}\Delta^{-}
              \end{array} \right)+\cdots\right\} \, , \\
 &=& \frac{1-\left(
          \begin{array}{cc}
            0  &  G_0^{+}\Delta^{-} \\
            G_0^{-}\Delta^{+}  & 0
              \end{array} \right)}{\left(
          \begin{array}{cc}
            \left[ G_0^{+} \right]^{-1}  & 0\\
            0  &  \left[ G_0^{-} \right]^{-1}
              \end{array}\right)\left\{ 1-\left(
          \begin{array}{cc}
            G_0^{+}\Delta^{-}G_0^{-}\Delta^{+}  & 0  \\
             0 & G_0^{-}\Delta^{+}G_0^{+}\Delta^{-}
              \end{array} \right)\right\} }\, ,  \nonumber \\
              &=& \frac{1-\left(
          \begin{array}{cc}
            0  &  G_0^{+}\Delta^{-} \\
            G_0^{-}\Delta^{+}  & 0
              \end{array} \right)}{\left(
          \begin{array}{cc}
            \left[ G_0^{+} \right]^{-1} - \Delta^{-}G_0^{-}\Delta^{+}& 0\\
            0  &  \left[ G_0^{-} \right]^{-1}-\Delta^{+}G_0^{+}\Delta^{-}
              \end{array}\right) } \, , \\
              &=&    \left(
          \begin{array}{cc}
            G^+  &  -G_0^{+}\Delta^{-}G^- \\
            -G_0^{-}\Delta^{+}G^+  & G^-
              \end{array} \right) = \left(
          \begin{array}{cc}
            G^+  &  \Xi^- \\
            \Xi^+  & G^-
              \end{array} \right) \,     ,
\end{eqnarray}
with
\begin{eqnarray}
G^+&=&\frac{1}{\left[G_0^{+} \right]^{-1} - \Delta^{-}G_0^{-}\Delta^{+}} \, , \\
G^-&=&\frac{1}{\left[G_0^{-}
\right]^{-1}-\Delta^{+}G_0^{+}\Delta^{-}} \, .
\end{eqnarray}
In the 2SC and g2SC phases with the basis $\Psi$ in the color-flavor
spaces
 (see Eq.(6)), the matrix $\Delta^+$ and $\Delta^-$ (see
Eqs.(8-10)) are skew diagonal, the following matrix is strictly
diagonal,
\begin{eqnarray}
\left(
          \begin{array}{cc}
            G_0^{+}\Delta^{-}G_0^{-}\Delta^{+}  & 0  \\
             0 & G_0^{-}\Delta^{+}G_0^{+}\Delta^{-}
              \end{array} \right) \, ,
              \end{eqnarray}
 the diagonal and shew-diagonal matrix elements decouple from each other in Eq.(16).
\footnote{The results in Eqs.(18-20) certainly can be derived by the
standard linear algebra manipulations, in this article, we propose
to separate the diagonal and shew-diagonal matrix elements in
Eq.(16) intuitively, the matrix in the denominator in Eq.(17) is
block diagonal, for the  2SC and g2SC states, the blocks
$G_0^{+}\Delta^{-}G_0^{-}\Delta^{+}$ and
$G_0^{-}\Delta^{+}G_0^{+}\Delta^{-}$ are strictly diagonal,  the
matrix can be inverted easily. The general strategy be, choose the
suitable basis for the matrixes $\Delta^+$ and $\Delta^-$ to
diagonalize or skew-diagonalize them in the color-flavor spaces, the
resulting blocks $G_0^{+}\Delta^{-}G_0^{-}\Delta^{+}$ and
$G_0^{-}\Delta^{+}G_0^{+}\Delta^{-}$ are strictly diagonal or mostly
diagonal with some non-diagonal blocks. The matrix $G^{-1}$ can also
be diagonalized by analytical plus numerical calculations, which is
performed in Ref.\cite{Ruster}. Furthermore, there are also other
works using the energy projective operators to invert the propagator
$G^{-1}$, for example, Ref.\cite{ForExample}. }
 We can read out the  matrix elements
              $G^{\pm}_{ru}$ , $G^{\pm}_{rd}$, $G^{\pm}_{gu}$ ,$G^{\pm}_{gd}$,
              $\Xi^{\pm}_{ru,gd}$ , $\Xi^{\pm}_{rd,gu}$, $\Xi^{\pm}_{gd,ru}$ , $\Xi^{\pm}_{gu,rd}$ from
              Eqs.(17-20)  straightforwardly with the help of the energy projective  operators.
For example,
\begin{eqnarray}
G^{\pm}_{ru}&=&\frac{1}{\left[G_{0}^{\pm} \right]^{-1}_{ru} -
\Delta(-i\gamma_5)\left[G_0^{\mp}\right]_{gd}\Delta (-i\gamma_5)} \, , \nonumber \\
& = & \frac{1}{ \gamma_0(p_0-E_{ru}^{\mp})\Lambda_{+} +
\gamma_0(p_0+E_{ru}^{\pm})\Lambda_{-}  + \Delta \gamma_5
(\frac{\gamma_0\tilde\Lambda_{+}}{p_0+E_{gd}^{\mp}} +
\frac{\gamma_0\tilde \Lambda_{-}}{p_0-E_{gd}^{\pm}}) \Delta \gamma_5} \, ,\nonumber \\
& = &
\frac{p_0-E_{gd}^{\pm}}{\left(p_0+{E_{ru}^{\pm}}\right)\left(p_0-{E_{gd}^{\pm}}\right)
-\Delta^2}\gamma_0\tilde\Lambda_{+} +
\frac{p_0+E_{gd}^{\mp}}{\left(p_0-{E_{ru}^{\mp}}\right)\left(p_0+{E_{gd}^{\mp}}\right)-\Delta^2}\gamma_0\tilde\Lambda_{-}\
,   \nonumber \\
G^{\pm}_{gd}& = &G^{\pm}_{ru} (ru \leftrightarrow gd )  \, ,
\nonumber
\\
\Xi^{\pm}_{ru,gd}&=&\frac{-i\Delta\gamma_5}{\left(p_0+{E_{ru}^{\pm}}\right)
\left(p_0-{E_{gd}^{\pm}}\right) -\Delta^2}\tilde\Lambda_{+}
+\frac{-i\Delta\gamma_5}{\left(p_0-{E_{ru}^{\mp}}\right)\left(p_0+{E_{gd}^{\mp}}\right)-\Delta^2}\tilde\Lambda_{-}\,
,
\nonumber \\
\Xi^{\pm}_{ru,gd}&=&\Xi^{\pm}_{ru,gd}(ru \leftrightarrow gd).
\end{eqnarray}
We can write down other matrix elements analogously with simple
substitution.

In the CFL and gCFL phases, the scalar diquark condensates  can be
parameterized as
\begin{equation}
\langle q^\alpha_i C\gamma_5 q^\beta_j \rangle \sim \Delta_1
\varepsilon^{\alpha\beta 1}\varepsilon_{ij1} \!+\! \Delta_2
\varepsilon^{\alpha\beta 2}\varepsilon_{ij2} \!+\! \Delta_3
\varepsilon^{\alpha\beta 3}\varepsilon_{ij3} \, ,
\end{equation}
 the condensates are  Lorentz scalars, antisymmetric in
Dirac indices, antisymmetric in color (with the strongest attraction
between quarks) and antisymmetric in flavor. The gap parameters
describe $d-s$, $s-u$ and $u-d$ Cooper pairs, respectively. We can
introduce the Nambu-Gorkov formulation  in the three-flavor case,
\begin{eqnarray}
\Psi^T&=&(q_{ru},q_{gd},q_{bs},q_{rd},q_{gu},q_{rs},q_{bu},q_{gs},q_{bd};q^C_{ru},q^C_{gd},q^C_{bs},q^C_{rd},q^C_{gu},q^C_{rs},q^C_{bu},q^C_{gs},q^C_{bd})
\, , \nonumber
\\
\bar{\Psi}&=&(\bar{q}_{ru},\bar{q}_{gd},\bar{q}_{bs},\bar{q}_{rd},\bar{q}_{gu},\bar{q}_{rs},\bar{q}_{bu},\bar{q}_{gs},\bar{q}_{bd};\bar{q}^C_{ru},\bar{q}^C_{gd},\bar{q}^C_{bs},\bar{q}^C_{rd},\bar{q}^C_{gu},\bar{q}^C_{rs},\bar{q}^C_{bu},\bar{q}^C_{gs},\bar{q}^C_{bd})\,
. \nonumber \\
\end{eqnarray}
 The inverse Nambu-Gorkov propagator is  a $18\times 18$ matrix in the color-flavor spaces,
\begin{eqnarray} \Delta^-&=& -i\gamma_5\left(
\begin{array}{ccccccccc}
 0 & \Delta_3 & \Delta_2 & 0 & 0 & 0 & 0 & 0 & 0 \\
 \Delta_3 & 0 & \Delta_1 & 0 & 0 & 0 & 0 & 0 & 0 \\
\Delta_2 & \Delta_1 & 0  & 0 & 0 & 0 & 0 & 0 & 0 \\
  0 & 0 & 0 & 0 & -\Delta_3 & 0 & 0 & 0 & 0 \\
  0 & 0 & 0 & -\Delta_3 & 0 & 0 & 0 & 0 & 0 \\
  0 & 0 & 0 & 0 & 0 & 0 & -\Delta_2 & 0 & 0 \\
  0 & 0 & 0 & 0 & 0 & -\Delta_2 & 0 & 0 & 0 \\
  0 & 0 & 0 & 0 & 0 & 0 & 0 & 0 & -\Delta_1 \\
  0 & 0 & 0 & 0 & 0 & 0 & 0 & -\Delta_1 & 0 \\
\end{array}
\right) \, , \\
&=&-i\gamma_5\left(
\begin{array}{cccc}
 A & 0& 0 & 0  \\
 0 & B &  0 & 0 \\
 0  & 0 & C & 0  \\
  0 & 0 & 0 & D  \\
\end{array}
\right) \, , \\
G_0^{\pm}&=& \left(
\begin{array}{ccccccccc}
G_{0ru}^{\pm} & 0 & 0 & 0 & 0 & 0 & 0 & 0 & 0 \\
 0 & G_{0gd}^{\pm} & 0 & 0 & 0 & 0 & 0 & 0 & 0 \\
0 & 0 & G_{0bs}^{\pm}  & 0 & 0 & 0 & 0 & 0 & 0 \\
  0 & 0 & 0 & G_{0rd}^{\pm} & 0 & 0 & 0 & 0 & 0 \\
  0 & 0 & 0 & 0 & G_{0gu}^{\pm} & 0 & 0 & 0 & 0 \\
  0 & 0 & 0 & 0 & 0 & G_{0rs}^{\pm}&0 & 0 & 0 \\
  0 & 0 & 0 & 0 & 0 & 0 & G_{0bu}^{\pm} & 0 & 0 \\
  0 & 0 & 0 & 0 & 0 & 0 & 0 & G_{0gs}^{\pm} & 0 \\
  0 & 0 & 0 & 0 & 0 & 0 & 0 & 0 & G_{0bd}^{\pm} \\
\end{array}
\right) \, ,
\end{eqnarray}
where $A$ is $3\times3$ block matrix, $B$, $C$ and $D$ are
$2\times2$ block matrixes. The matrix
\begin{eqnarray}
\left(
\begin{array}{cc}
G_0^{+}\Delta^{-}G_0^{-}\Delta^{+}&0  \\
 0 & G_0^{-}\Delta^{+}G_0^{+}\Delta^{-} \\
\end{array}
\right)
\end{eqnarray}
is block diagonal,  the matrix $G_0^{+}\Delta^{-}G_0^{-}\Delta^{+}$
and $G_0^{-}\Delta^{+}G_0^{+}\Delta^{-}$ are diagonal for the six
basis $ (rd, gu, rs, bu, gs, bd)$ and have  non-diagonal blocks
concerning the matrix $A$ for the three basis $(ru,gd,bs)$.  The
diagonal matrix elements of   the inverse matrix of
\begin{eqnarray}
\left(
\begin{array}{cc}
\left[G_0^{+}\right]^{-1}-\Delta^{-}G_0^{-}\Delta^{+}&0  \\
 0 & \left[G_0^{-}\right]^{-1}-\Delta^{+}G_0^{+}\Delta^{-} \\
\end{array}
\right)
\end{eqnarray}
are the inverse of the corresponding   matrix elements. We can read
out the matrix elements $G^{\pm}_{rd}$
,$G^{\pm}_{gu}$,$G^{\pm}_{rs}$,$G^{\pm}_{bu}$,$G^{\pm}_{gs}$,
$G^{\pm}_{bd}$ directly from the Eqs.(17-20) with the help of the
energy  projective operators. For example,
\begin{eqnarray}
G^{\pm}_{rd}&=&\frac{1}{\left[G_{0}^{\pm} \right]^{-1}_{rd} -
\Delta_3(-i\gamma_5)\left[G_0^{\mp}\right]_{gu}\Delta_3 (-i\gamma_5)} \nonumber \\
& = &
\frac{p_0-E_{gu}^{\pm}}{\left(p_0+{E_{rd}^{\pm}}\right)\left(p_0-{E_{gu}^{\pm}}\right)
-\Delta_3^2}\gamma_0\tilde\Lambda_{+} +
\frac{p_0+E_{gu}^{\mp}}{\left(p_0-{E_{rd}^{\mp}}\right)\left(p_0+{E_{gu}^{\mp}}\right)-\Delta_3^2}\gamma_0\tilde\Lambda_{-}\
, \nonumber \\
G^{\pm}_{gu}& = &G^{\pm}_{gu}(gu \leftrightarrow rd) \, , \nonumber
\\
\Xi^{\pm}_{rd,gu}&=&\frac{-i\Delta_3\gamma_5}{\left(p_0+{E_{rd}^{\pm}}\right)\left(p_0-{E_{gu}^{\pm}}\right)
-\Delta_3^2}\tilde\Lambda_{+}
+\frac{-i\Delta_3\gamma_5}{\left(p_0-{E_{rd}^{\mp}}\right)\left(p_0+{E_{gu}^{\mp}}\right)-\Delta_3^2}\tilde\Lambda_{-}
\, , \nonumber \\
\Xi^{\pm}_{gu,rd}&=&\Xi^{\pm}_{rd,gu}(rd \leftrightarrow gu) \, .
\end{eqnarray}
We can write down the corresponding ones for the $G_{rs}^{\pm}$,
  $G_{bu}^{\pm}$, $ G_{gs}^{\pm} $, $G_{bd}^{\pm}$, $
  \Xi^{\pm}_{rs,bu}$, $ \Xi^{\pm}_{bu,rs}$, $ \Xi^{\pm}_{gs,bd}$, $
  \Xi^{\pm}_{bd,gs}$ analogously.

For the non-diagonal blocks concerning the $A$ in Eqs.(28-29), in
general one may expect
 a linear transformation for the basis $(ru,gd,bs)$ can result in
 diagonal matrixes for both the block $A$ and the corresponding
 $\left[G^{\pm}_0\right]^{-1}$, however, it is not the case.
The color superconducting phases  may exist in the core of compact
stars, where the bulk matter should satisfy the charge neutrality
condition and the $\beta$-equilibrium.  The gravitation force is
much weaker than the electromagnetic and the color forces,  any
electric charges or color charges can forbid the formation of bulk
matter. For a neutral two-flavor color superconductor, the quark
chemical potentials in color and flavor spaces can be expressed in
terms of baryon chemical potential $\mu$, electrical chemical
potential $\mu_e$, and color chemical potentials $\mu_3$ and
$\mu_8$,
\begin{eqnarray}
\mu_{\alpha  i, \beta j}&=&\mu \delta_{\alpha\beta}\delta_{ij}
-\mu_e Q_{ij}\delta_{\alpha\beta} +\mu_3
T^3_{\alpha\beta}\delta_{ij}+\frac{2}{\sqrt{3}}\mu_8
T^8_{\alpha\beta}\delta_{ij} \, ,
\end{eqnarray}
 while for the three-flavor quark system, the chemical potentials are
 taken as
 \begin{eqnarray}
\mu_{\alpha  i, \beta j}&=&\mu \delta_{\alpha\beta}\delta_{ij}
-\mu_e Q_{ij}\delta_{\alpha\beta} +\mu_3
T^3_{\alpha\beta}\delta_{ij}+\frac{2}{\sqrt{3}}\mu_8
T^8_{\alpha\beta}\delta_{ij} -\frac{\delta m^2}{2\mu}\delta_{i 3}
\delta_{j 3}\delta_{\alpha\beta} \, .
\end{eqnarray}
From above equations for the chemical potentials, we can see that
they are far from being equal. The charge neutrality condition puts
strong constraints on both the two-flavor and three-flavor systems
and results  in the g2SC and gCFL phases.  In the gCFL phase, the up
block $A$ for the matrix $\Delta^-$ in Eq.(25) and the corresponding
one for $\left[G^{\pm}_0\right]^{-1}$ in Eq.(27) can not be
diagonalized  with the same basis. Fortunately, we can solve the SDE
for the gaps $\Delta_1$, $\Delta_2$, $\Delta_3$ with the six
 basis $ (rd, gu, rs, bu, gs, bd)$.

 The explicit and simple expressions  for the quark propagators  are of great
importance in solving the gap equations  and determining  the gap
parameters $\Delta$, $\Delta_1$, $\Delta_2$,  $\Delta_3$.
\begin{eqnarray}
 G^{-1}(p) - G^{-1}_0 (p) ~=~ i g^2 \int {d^4k \over (2
\pi)^4}
  ~\Gamma^A_\mu
~G(k)~ \Gamma^B_\nu  ~D^{\mu \nu}_{AB} (k-p)~, \end{eqnarray} where
\begin{eqnarray} \label{G} \Gamma^A_\mu~=~ \left( \begin{array}{cc}
\gamma_\mu T^A  & 0 \\
0 & C (\gamma_\mu T^A)^T C^{-1}
\end{array} \right)~.
\end{eqnarray}
The  gluon propagator $D^{\mu \nu}_{AB}$ has complex energy and
chemical potentials dependence, include the effects of Landau
damping and Debye screening induced by the medium \cite{Bellac}.
Furthermore, the contributions come from the scalar diquark
condensates should also be taken into account in the color
superconducting phases. The gluon self-energy has been investigated
in the  g2SC  and gCFL phases. The results show that the gauge
bosons connected with the broken generators  have imaginary Meissner
screening masses \cite{GaplessMass}, the chromomagnetic instability
should be taken into account in  a consistent analysis.  The
explicit expressions for the quark propagators  can greatly
facilitate the calculations, especially in performing the Matsubara
frequency summation. At low baryon density, the coupling constant
$g^2$ can not be factorized out, $g^2D^{\mu \nu}_{AB}$ should be
approximated by some phenomenologically potentials just as the
conventional SDE  at zero baryon density \cite{SDE}.

In the  CFL state with the same chemical potential (or in the limit
$\mu\rightarrow \infty$) and degenerate mass $m$ ($\delta m
\rightarrow 0$),  we can perform a linear transformation to
diagonalize the matrix $A$ and
$diag(G^{\pm}_{0ru},G^{\pm}_{0gd},G^{\pm}_{0bs} )$ simultaneously,
\begin{eqnarray}\left(
\begin{array}{c}
 ru   \\
 gd \\
 bs  \\
  \end{array}
\right) &=&\left(
\begin{array}{cccc}
 \frac{1}{\sqrt{3}} & \frac{2}{\sqrt{6}}& 0   \\
 \frac{1}{\sqrt{3}} & -\frac{1}{\sqrt{6}} &  \frac{1}{\sqrt{2}}  \\
 \frac{1}{\sqrt{3}}  & -\frac{1}{\sqrt{6}} & -\frac{1}{\sqrt{2}}  \\
  \end{array}
\right)\left(
\begin{array}{c}
 x_1  \\
 x_2 \\
 x_3  \\
  \end{array}
\right) \\
\end{eqnarray}
With the new basis $(x_1,x_2,x_3)$, the matrix $A$ takes the
following form
\begin{eqnarray}\left(
\begin{array}{ccc}
 2 \Delta & 0& 0   \\
 0 & -\Delta &  0  \\
 0  & 0 & -\Delta  \\
  \end{array}
\right) \, .
\end{eqnarray}
The matrix  $G_0^{+}\Delta^{-}G_0^{-}\Delta^{+}$  and
$G_0^{-}\Delta^{+}G_0^{+}\Delta^{-}$ are strictly diagonal for the
nine basis $ (x_1,x_2,x_3,rd, gu, rs, bu, gs, bd)$, we can read out
the $G^{\pm}$ for the basis $(x_1,x_2,x_3)$ from Eqs.(17-20)
directly.

\begin{eqnarray}
G^{\pm}_{1}&=&\frac{1}{\left[G_{0}^{\pm} \right]^{-1}_{1} -
2\Delta(-i\gamma_5)\left[G_0^{\mp}\right]_{1}2\Delta (-i\gamma_5)} \, , \nonumber \\
& = & \frac{p_0-E_{1}^{\pm}}{p_0^2-\left({E_{1}^{\pm}}\right)^2
-4\Delta^2}\gamma_0\tilde\Lambda_{+} +
\frac{p_0+E_{1}^{\mp}}{p_0^2-\left({E_{1}^{\mp}}\right)^2
-4\Delta^2}\gamma_0\tilde\Lambda_{-}\
, \nonumber \\
G^{\pm}_{2}& = &G^{\pm}_{3}=G^{\pm}_1 (2\Delta \rightarrow \Delta )
\,  , \nonumber
\\
\Xi^{\pm}_{1,1}&=&\frac{-i2\Delta\gamma_5}{p_0^2-\left({E_{1}^{\pm}}\right)^2
-4\Delta^2}\tilde\Lambda_{+}
+\frac{-i2\Delta\gamma_5}{p_0^2-\left({E_{1}^{\mp}}\right)^2
-4\Delta^2}\tilde\Lambda_{-}\, ,
\nonumber \\
\Xi^{\pm}_{2,2}&=&\Xi^{\pm}_{3,3}=
\Xi^{\pm}_{1,1}(2\Delta\rightarrow\Delta) \, .
\end{eqnarray}
The block matrix for the $G$ and $\Xi$ with the basis $(ru,gd,bs)$
can be written as
\begin{eqnarray}\frac{1}{3}\left(
\begin{array}{ccc}
 G_1^{\pm}+2G_2^{\pm} & G_1^{\pm}-G_2^{\pm}& G_1^{\pm}-G_2^{\pm}   \\
 G_1^{\pm}-G_2^{\pm} & G_1^{\pm}+2G_2^{\pm} &  G_1^{\pm}-G_2^{\pm}  \\
 G_1^{\pm}-G_2^{\pm}  & G_1^{\pm}-G_2^{\pm} & G_1^{\pm}+2G_2^{\pm}  \\
  \end{array}
\right) \, ,
\end{eqnarray}
\begin{eqnarray}\frac{1}{3}\left(
\begin{array}{ccc}
 \Xi^{\pm}_1+2\Xi^{\pm}_2 & \Xi^{\pm}_1-\Xi^{\pm}_2& \Xi^{\pm}_1-\Xi^{\pm}_2   \\
 \Xi^{\pm}_1-\Xi^{\pm}_2 & \Xi^{\pm}_1+2\Xi^{\pm}_2 &  \Xi^{\pm}_1-\Xi^{\pm}_2  \\
 \Xi^{\pm}_1-\Xi^{\pm}_2  & \Xi^{\pm}_1-\Xi^{\pm}_2 & \Xi^{\pm}_1+2\Xi^{\pm}_2  \\
  \end{array}
\right) \, .
\end{eqnarray}
From above matrixes, we can see that for the normal propagators, due
the  presence of the scalar diquark condensates,  all the matrix
elements $\langle q_{\alpha i} \bar{q}_{\beta j}\rangle$ are
non-zero, while only the diagonal elements in the color-flavor
spaces are non-zero in the ordinary vacuum.

\section{Conclusion }
In this article, we propose a new approach to obtain the
Nambu-Gorkov propagator intuitively with some linear algebra
techniques in presence of the scalar diquark condensates. With the
help of energy projective operators, we  obtain relatively simple
expressions for the quark propagators, which can greatly facilitate
the calculations  in solving the SDE to obtain the gap parameters.

\section*{Acknowledgment}
This  work is supported by National Natural Science Foundation,
Grant Number 10405009,  and Key Program Foundation of NCEPU. The
authors are indebted to Dr. J.He (IHEP), Dr. X.B.Huang (PKU) and Dr.
L.Li (GSCAS) for numerous help, without them, the work would not be
finished.

\end{document}